\documentstyle[aps,amssymb,epsfig,twocolumn]{revtex}  
\begin{document}
\textheight 241mm
\textwidth 182mm
\wideabs
{
\title{\LARGE 
{\sf
Xe films on a decagonal Al-Ni-Co quasicrystal surface}}
\author{
Wahyu Setyawan$^{1}$, Nicola Ferralis$^{2}$, Renee D. Diehl$^{2}$, Milton W. Cole$^{2}$, Stefano Curtarolo$^{1,3}$}
\address{
$^1$Department of Mechanical Engineering and Materials Science, Duke University, Durham, NC 27708 \\
$^2$Department of Physics and Materials Research Institute, Penn State University, University Park, PA 16802\\
$^3${corresponding author, e-mail: stefano@duke.edu}
}
\date{\today}
\maketitle
\begin{abstract}
The grand canonical Monte Carlo method is employed to study the
adsorption of Xe on a quasicrystalline Al-Ni-Co surface. The
calculation uses a semiempirical gas-surface interaction, based on
conventional combining rules and the usual Lennard-Jones Xe-Xe
interaction. The resulting adsorption isotherms and calculated
structures are consistent with the results of LEED experimental data.
In this paper we focus on five features not discussed earlier
(Phys. Rev. Lett. {\bf 95}, 136104 (2005)): the range of the average
density of the adsorbate, the order of the transition, the
orientational degeneracy of the ground state, the isosteric heat of
adsorption of the system, and the effect of the vertical cell
dimension.
\end{abstract}
%\pacs{PACS numbers: 68.43.-h, 61.44.Br, 68.55.Ac}
}

%%%%%%%%%%%%%%%%%%%%%%%%%%%%%%%%%%%%%%%%%%%%%%%%%%%%%%%%%%%%%%%%%%%%%%%%%%%%%%%%
%%%%%%%%%%%%%%%%%%%%%%%%%%%%%%%%%%%%%%%%%%%%%%%%%%%%%%%%%%%%%%%%%%%%%%%%%%%%%%%%
%%%%%%%%%%%%%%%%%%%%%%%%%%%%%%%%%%%%%%%%%%%%%%%%%%%%%%%%%%%%%%%%%%%%%%%%%%%%%%%%

\section{Introduction}

The adsorption of simple gases on essentially flat surfaces has been
studied extensively for the last 50 years, beginning with early
experiments on exfoliated graphite\cite{ref1,ref2,ref3}. Many
phenomena that are found to occur in these systems are reasonably well
described by the two-dimensional (2D)
approximation\cite{ref3}. However, this approximation fails to
capture some of the more intriguing kinds of behavior, which make the
subject of monolayer films diverse and challenging. A variety of
monolayer phenomena can be attributed to competing adsorbate-adsorbate
and adsorbate-substrate interactions; the relevant variables are the
substrate symmetry and the interaction strengths and length scales of
the adsorbate relative to those of the substrate\cite{ref10,ref11}.
 
Recently, our group has begun to explore the behavior of simple gases
physisorbed on quasicrystalline surfaces\cite{ref12,ref13}. The first
system that we have explored extensively is Xe on the surface of the
quasicrystal d-Al$_{73}$Ni$_{10}$Co$_{17}$\cite{ref14}. Ours is not
the first study of film growth on quasicrystalline surfaces. A variety
of growth modes has been seen in previous studies on a variety
of substrates\cite{ref15}.  A key difference between our work
and the previous investigations is that Xe is expected to be
physically adsorbed, although we will see that the atomic binding
energy lies near the upper limit of the regime normally ascribed to
physical interactions (binding energy $\approx$ 0.3 eV). One
advantage of physisorption relative to stronger binding adsorption is
the presumption that the physisorption process does not alter the
interacting partners significantly.  That is, the surfaces do not
reconstruct due to the film's presence and the adatoms retain their
chemical identity, interacting with each other with forces similar to
their gas phase interaction. A substantial body of experimental data
for a variety of adsorption systems has provided strong support for
this point of view\cite{ref3}.

In this paper we focus on five features not discussed earlier
\cite{ref12}: the range of the average density of the adsorbate, the
order of the transition, the orientational degeneracy of the ground
state, the isosteric heat of adsorption of the system, and the effect
of the vertical cell dimension.  In the following section, we discuss
the methods used in this simulation study.  Section II introduces
briefly the method.  Section III reports the results and compares them
with recent experiments from our group.  Section IV summarizes these
results, draws conclusions and comments on strategies for future
research in this area.

%%%%%%%%%%%%%%%%%%%%%%%%%%%%%%%%%%%%%%%%%%%%%%%%%%%%%%%%%%%%%%%%%%%%%%%%%%%%%%
\section{Method}
 
We study the adsorption of Xe on the tenfold surface of the
decagonal Al$_{73}$Ni$_{10}$Co$_{17}$ quasicrystal (QC) using the
grand canonical Monte Carlo simulation method
(GCMC)\cite{ref16,ref17}.  Since this technique is widely used and we
have described our method in detail previously\cite{ref13,ref18}, only
a brief description is given here. At constant volume, $V$, and
temperature, $T$, the GCMC method explores the configurational phase
space, using the familiar Metropolis algorithm, and finds the
equilibrium number of adsorbed atoms, $N$, as a function of the
chemical potential, $\mu$, of Xe.  $\mu$ is related to the pressure of
the coexisting gas, which is taken to be ideal. 
We also determine density profiles, $\rho(x,y)$, 
and adsorption isotherms, $\rho_N$, as a function of the pressure,
$P(T, \mu)$.  For each data point in an isotherm, 18 million GCMC steps
(each step being an attempted creation, deletion, or displacement of
an atom) are performed to reach nominal equilibrium and 27 million
steps are performed in the subsequent data-gathering phase.
Displacements, creations, and destructions of atoms are executed with
probabilities equal to 0.2, 0.4, and 0.4, respectively\cite{ref18}.

The interaction potentials used in the calculations are based on the
12-6 Lennard-Jones (LJ) functional form of pair interaction.  The
construction of the interactions is described in detail in
references\cite{ref12,ref19,ref20,ref21}.  Using these potentials, we
perform simulations in a tetragonal cell.  The height of the cell,
along the $z$ (surface-normal) direction, is chosen to be 10 nm (long
enough to contain $\sim$20 layers of Xe).  At the top of the cell, a
hard-wall reflective potential is simulated to confine the vapor. The
base of the cell has dimensions of 5.12 x 5.12 nm$^2$ and periodic
boundary conditions are employed along the $x$ and $y$
directions. These periodic boundary conditions render the surface
effectively infinite; we use a relatively large cutoff
($5\sigma_{gg}$) to minimize long range interaction corrections.

Figure \ref{fig1}(a) shows the
function $V_{min}(x,y)$, which is calculated by minimizing the
adsorption potential along the $z$ direction at every value of the
planar coordinates $(x,y)$:
\begin{equation} 
	\left. V_{min}(x,y) \equiv min\left(V(x,y,z)\right)\right|_{along\,\,z}.
\end{equation}
Such a figure reveals the fivefold rotational symmetry of the substrate. 
Dark regions in the figure correspond to the most attractive regions 
on the substrate. By choosing appropriate sets of five dark spots, 
we can identify pentagons, which correspond to the inflated tile of a 
pentagonal Penrose tiling. The sizes of the pentagons follow the 
inflationary property of the QC structure. 

%%%%%%%%%%%%%%%%%%%%%%%%%%%%%%%%%%%%%%%%%%%%%%%%%%%%%%%%%%%%%%%%%%%%%%%%%%%%%%%%
\section{Results and Discussion}

\begin{figure}[htb]
  \centerline{\epsfig{file=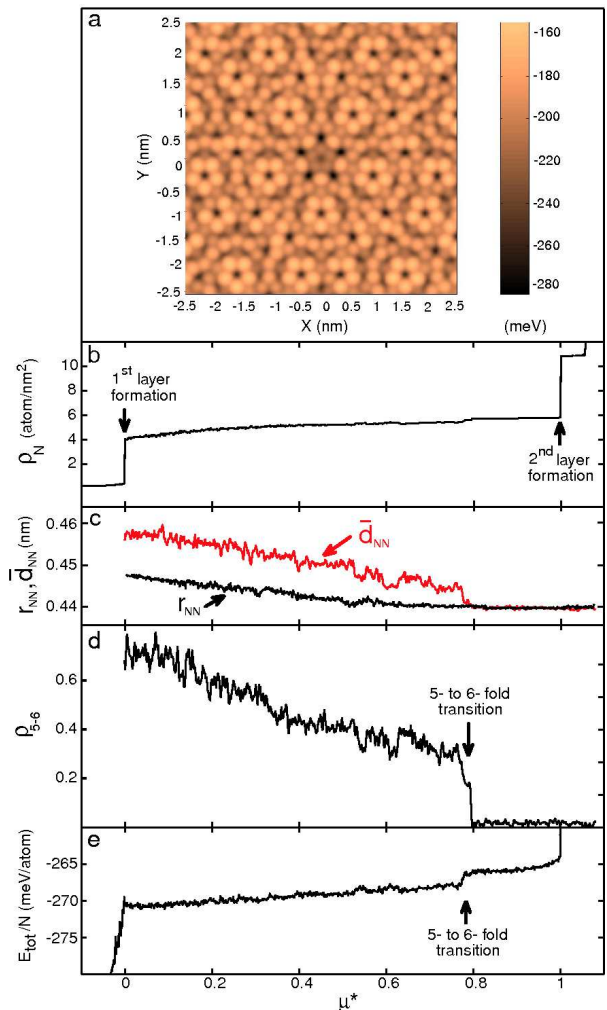,width=80mm,clip=}}
  \caption{\small (color online).  
    (a) Minimum potential energy surface,
    $V_{min}(x.y)$, for Xe on a 5.12 x 5.12 nm$^{2}$ section of QC. The scale at the
    right shows the energy scale, which ranges from -280 to -160
    meV. 
    (b) Adsorption isotherm, $\rho_N$, versus the reduced chemical potential, $\mu^\star$, at $T$ = 77K.
    (c) Nearest neighbor distance defined from the pair correlation function, $r_{NN}$, (black line), and
    average spacing between neighbors at equilibrium, $\bar{d}_{NN}$, (red line). 
    (d) Order parameter $\rho$$_{5-6}$ (probability of fivefold defects, defined in Eq. \ref{eq3})
    versus the reduced chemical potential, $\mu^\star$, at $T$ = 77K.
    (e) Total energy per atom at $T$ = 77K.
    The transition, which is defined as the point in $\mu^\star$ above which the order parameter
    remains nearly constant, occurs at $\mu^\star_{tr}\sim$0.8.
    The discontinuity in $E_{tot}/N$ around $\mu^\star_{tr}\sim$ 0.8 
    indicates a latent heat of the transition.
    The order parameter $\rho$$_{5-6}$ after the transition is $\sim$ 0.017.
    }
  \label{fig1}
\end{figure}

A layer-by-layer growth mode for Xe on this QC surface was observed
and reported earlier\cite{ref12}.  Figure \ref{fig1}(b) shows a
particular isotherm at $T$=77K with the formation of the first and
second layers.

%%%%%%%%%%%%%%% DENSITY OF THE ADSORBATE
{\it Density of the adsorbate.} A detailed study of the density
profiles of the isotherms shows that the range of the density of the
monolayer is considerably larger than that of the bilayer. For
example, at 77K, the average density of the monolayer increases from
4.09 atoms/nm$^2$ at its formation to 5.74 atoms/nm$^2$ at its
completion (a 40\% increase) whereas for the bilayer, it increases
from 10.84 atoms/nm$^2$ at its formation to 10.98 atoms/nm$^2$ at its
completion (less than 2\% increase). The density increase of the
monolayer is several times larger than that observed experimentally
for Xe on Ag(111), a much flatter substrate\cite{newref1}. The
difference is due to the much larger lateral variation in adsorption
energy experienced by the Xe on the QC surface.  The magnitude of this
variation is apparently much smaller for the second layer.

%%%%%%%%%%%%%%% ORDER OF THE TRANSITION
{\it Order of the transition.} An interesting phenomenon that we have
found in this system is the continuous rearrangement of Xe atoms in
the monolayer, which leads to the ordering transition from fivefold to
sixfold.  
Such reordering appears to be continuous in the evolution of the
location of the first peak in the pair correlation function shown in
Figure 2(b) of reference \cite{ref12}.
However, there are some peculiarities that must be explained in detail. 

To better characterize the evolution of the adsorption process
we define a reduced chemical potential $\mu^\star$, as:
	\begin{equation}
		\mu^\star \equiv \frac{\mu-\mu_1}{\mu_2-\mu_1},
	\end{equation}
where $\mu_1$ and $\mu_2$ are the chemical potentials at the onset of
the first and second layer formation, respectively. 
In addition, we introduce an order parameter, $\rho$$_{5-6}$, 
which is defined as the probability of fivefold defects:
\begin{equation}
  \rho_{5-6} \equiv \frac{N_5}{N_5 + N_6},
  \label{eq3}
\end{equation}
where $N$$_5$ and $N$$_6$ are the number of atoms having 2D
coordination number equal to 5 and 6, respectively. The 2D
coordination number is the number of neighboring atoms within a cutoff
radius of $1.366\cdot0.44$ nm = 0.601 nm 
($1.366=\cos(\pi/6)+1/2$ is the average of the 1st NN and the 2nd NN in a triangular lattice);
0.44 nm is taken from the first NN distance of Xe at 77K
(note that this distance does not change appreciably between 0.440 nm at 77K and 0.443 nm at 140K \cite{ref12}).
 
%
%\vspace{-2mm}

Figure \ref{fig1}(d) shows a plot of order parameter
$\rho$$_{5-6}$ versus the reduced chemical potential for $T=77$K. It
starts from $\sim$ 0.8 at the first layer formation and
gradually decreases. It drops and remains at a nearly constant value of
$\sim$ 0.017 for $\mu^\star \geq \mu^\star_{tr}\sim$ 0.8,
where $\mu^\star_{tr}$ is a transition reduced chemical potential at
77K. Figure \ref{fig1}(e) shows the total energy per atom,
$E_{tot}/N$. At the transition point $\mu^\star_{tr}$, the energy per
atom has a little step indicating a latent heat of the transition.
The discontinuity of the order parameter $\rho$$_{5-6}$ and the presence of
latent heat indicate that the transition is first-order.

In spite of this evidence for a first-order transition, the nearest neighbor
distance, labeled $r_{NN}$ in Figure \ref{fig1}(c), appears to change continuously.  
This nearest neighbor distance, also reported in our earlier paper \cite{ref12}, was
defined as the location of the first peak in the pair correlation function 
because the latter property is more directly comparable to diffraction measurements. 
Here we have also calculated the average spacing between neighbors, 
$\bar{d}_{NN}$, which is a thermodynamically meaningful quantity (related to the density).
This has a small discontinuity at the transition, providing additional evidence
for the first-order character of the transition. 
Both quantities, $r_{NN}$ and $\bar{d}_{NN}$, are shown in Figure \ref{fig1}(c). 
The NN Xe-Xe distance $r_{NN}$
decreases continuously as $P$ increases, starting from 0.45 nm and
saturating at 0.44 nm. The Xe-Xe distance reaches saturation value
before the appearance of the second layer; therefore, the transition
is complete within the first layer. We note that a similar decrease
in NN distance was measured for Xe/Ag(111), but in that case, the NN
spacing did not saturate before the onset of the second layer
adsorption\cite{newref1,newref7}. 

\begin{figure}[htb]
  \centerline{\epsfig{file=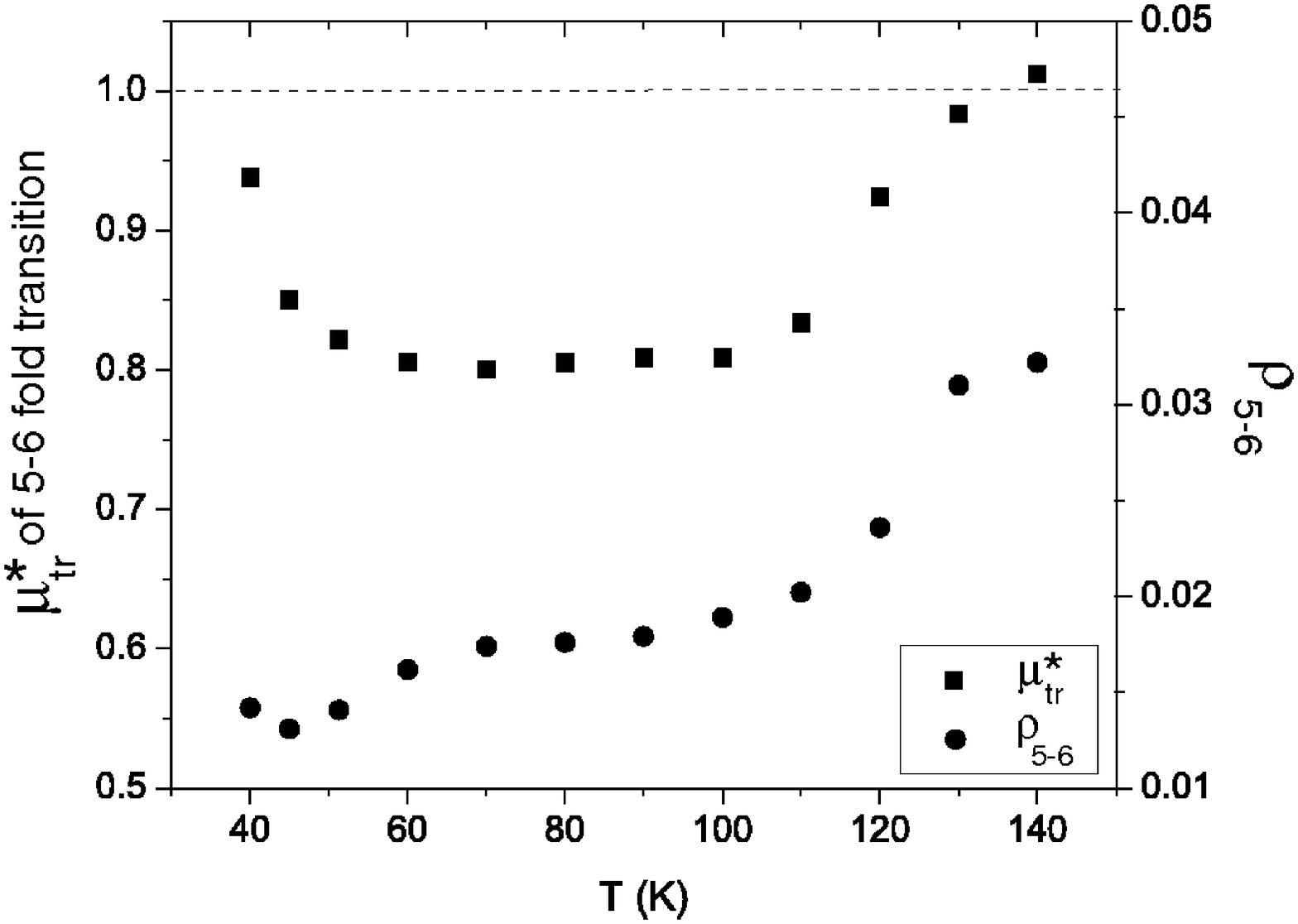,width=87mm,clip=}}
  \vspace{-1mm}
  \caption{\small
    Values of $\mu^\star_{tr}$ for the fivefold to sixfold 
    transition points from 40K to 140K (left axis). 
    Transition points at $\mu^\star_{tr}>1$ indicate 
    that a transfer of atoms from the second layer to the first layer is 
    required to complete the transition. 
    Also shown is the defect probability
    as a function of $T$ after the transition occurs (right axis), 
    indicating an increase in defect probability with $T$.}
  \label{fig2}
\end{figure}

Defects are present at all temperatures that we have simulated (in the
interval 20K to 286K). The probability of defects increases with
temperature, implying that their origin is entropic, as is the
case for a periodic crystal. Figure \ref{fig2}-(right axis) shows
that the defect probability increases as T increases, while Figure
\ref{fig2}-(left axis) shows the trend of the transition point in
function of T. At low temperatures, the sixfold ordering occurs
earlier (at lower $\mu^\star_{tr}$) as the temperature is increased
from 40K to 70K. This trend is expected because the ordering effect
imposed by the substrate corrugation becomes relatively smaller as the
temperature increases. However, this trend is not observed in the higher
temperature region (from 70K to 140K). In fact, at higher
temperatures the transition point shifts again to higher
$\mu^\star_{tr}$. This is most likely due to the monolayer becoming
less two-dimensional, allowing more structural freedom of the Xe atoms
and thus decreasing the effect of the repulsive Xe-Xe interaction that
would stabilize the sixfold structure. Transitions having critical
$\mu^\star_{tr}>1$ indicate that the onset of second-layer adsorption
occurs earlier than the transition to the sixfold structure. When the
second layer adsorbs at $T>$130K, the density of the monolayer
increases by a few percent, thereby increasing the effect of the
repulsive interactions and driving the fivefold to sixfold transition.

The fact that the structural transition for Xe occurs entirely within the
first layer suggests that atoms in subsequent layers are arranged in a
triangular lattice, which is indeed the case\cite{ref12}.
Furthermore, this finding suggests that the effect of the corrugation
of the adsorption potential on the structure of further layers is
quite small. The large corrugation experienced by the first layer
combined with the structural mismatch of the substrate and the Xe(111)
plane causes more variation in the local structures of the atoms in the
first layer. This explains why the first-layer step in the isotherm is
broader than the second-layer step, and why the density variation
during the evolution of the bilayer is smaller, as pointed out
earlier. When the second layer begins to form, our simulations
indicate a slightly (about 0.01 nm) larger NN spacing for the second
layer than for the first layer. A similar coexistence of two lattice
spacings was also found in calculations for Xe/Ag(111)\cite{newref2}
but was not observed experimentally for that system or for this one.

%\vspace{-6mm}
%\vspace{-2mm}

The observed transition from fivefold to sixfold order within the
first layer can be viewed as a commensurate-incommensurate transition
(CIT), since at the lower coverage, the layer is commensurate with the
substrate symmetry and aperiodic, while at higher coverage, it is
incommensurate with the substrate.  Such transitions within the first
layer have been observed before for adsorbed gases, perhaps most
notably for Kr on graphite\cite{ref25}.  There, as here for Xe, the Kr
forms a commensurate structure at low coverage, which is compressed
into an incommensurate structure at higher coverage.  The opposite
occurs for Xe on graphite, which is incommensurate at low coverage and
commensurate at high coverage\cite{ref26}. Such
commensurate-incommensurate transitions have been studied
theoretically in many ways, but perhaps most simply as a harmonic
system (balls and springs) having a natural spacing that experiences a
force field having a different spacing\cite{cit3}.  Such a transition
has been found to be first-order for strongly corrugated potentials
(in 1D) but continuous for more weakly corrugated
potentials\cite{ref27}.  The transition observed in our quasicrystal
surface suggests that system is within the regime of ``strong''
corrugation, which was not the case of Kr over graphite\cite{ref25}.
In fact, for the latter system, both commensurate and incommensurate
structures have sixfold symmetry.  A more relevant comparison may be
the transition of Xe on Pt(111), from a rectangular symmetry
incommensurate phase to a hexagonal symmetry commensurate one,
although in that case, the low-temperature phase was incommensurate.
That transition was also found to be continuous\cite{ref28}.
Therefore, while our simulations indicate that Xe on Al-Ni-Co
undergoes a CIT, as observed for other adsorbed gases, the observation
of a first-order CIT is new, to our knowledge, and likely arises from
the large corrugation. Simulations with other noble gases,
possessing different values of $\sigma_{gg}$ and $\epsilon_{gg}$,
would give insight into the origin and location of the
transition\cite{refprc}. Simulations carried out with decreasing
pressure in this region show some hysteresis. While hysteresis is
often interpreted as evidence of a first-order transition, this is not
necessarily the rule.  For example previous calculations have
exhibited hysteretic behavior within a monolayer on a very
heterogeneous surface, where no transition
occurs\cite{refnote1a,refnote1b}.

%%%%%%%%%%%%%%% ORIENTATIONAL DEGENERACY OF THE GROUND STATE
{\it Orientational degeneracy of the ground state.} In our earlier
paper, it was described that after the ordering transition is
complete, the resulting sixfold structure is aligned parallel to one
of the sides of the pentagons in the $V_{min}$ map of the adsorption
potential (there are five possible orientations).  In the
experiments, all five orientations are observed, due to the presence
of all possible alignments of hexagons along five sides of a pentagon
in the QC sample within the width of the electrons beam ($\sim$0.25
mm). In an ideal infinite GCMC framework the ground state of the
system would be degenerate and all five orientations would have the
same energy and be equally probable.  However, the square periodic
boundary conditions of our GCMC break this orientational degeneracy,
causing some orientations to become more likely to appear.

\begin{figure}[htb]
  \centerline{\epsfig{file=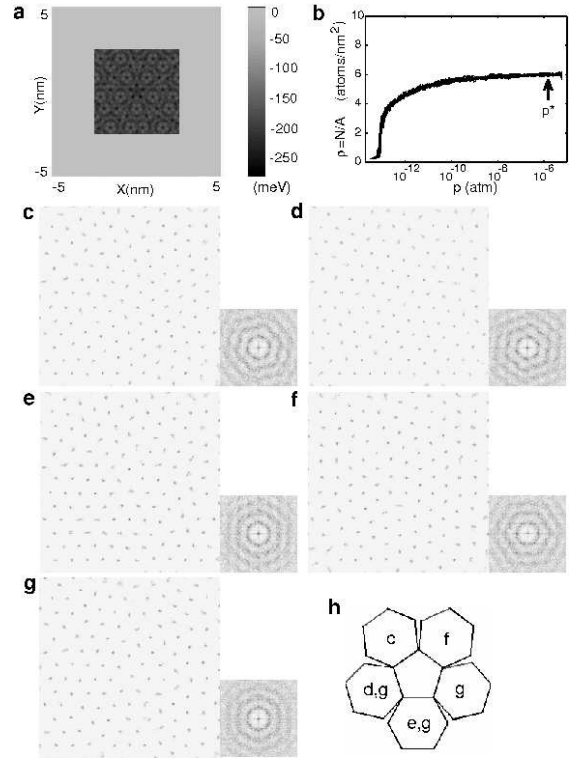,width=74mm,clip=}}
  \vspace{1mm}
  \caption{\small (a) Minimum potential energy surface of the
    adsorption potential with free boundary conditions. (b) Adsorption
    isotherms of the first layer from a set of 30 simulations at 77K
    using the free cell described in the paper. Five density profiles
    and FTs at point $p^\star$ of (b) are shown in (c) to (g),
    representing all possible orientations of hexagonal domains. (h)
    Schematic diagram illustrating the correspondence between the
    orientations of the hexagonal domains observed in the density
    profiles (c) to (g).}
  \label{fig3}
\end{figure}

To find all the possible orientations, we performed simulations with a
cell having free boundary conditions.  The cell is a 5.12 x 5.12
nm$^2$ quasicrystal surface surrounded by vacuum.  Figure \ref{fig3}(a)
shows the $V_{min}$ map of the adsorption potential.  Thirty
simulations at 77K are performed with this cell.  The isotherms from
these runs are plotted in Figure \ref{fig3}(b).  Only the first layer is
shown, and the finite size of the surface makes the growth of the
first layer continuous.  The density profiles $\rho(x,y)$ of all the
simulations are analyzed at point $p^\star$ of Figure \ref{fig3}(b).  In this cell,
all five orientations of hexagons are observed with equal frequency
indicating the orientational degeneracy of the ground state.  To represent the five
orientations, density profiles of five calculations (c, d, e, f, and
g) are shown in Figures \ref{fig3}(c) to \ref{fig3}(g) with their FT
plotted on the side.  Figure \ref{fig3}(h) presents a schematic
depiction of which orientations of hexagons are exemplified in each
simulation.

\begin{figure}[htb]
  \centerline{\epsfig{file=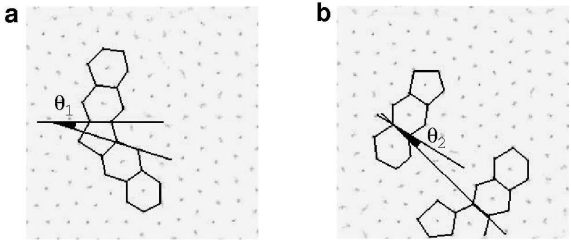,width=75mm,clip=}}
  \vspace{1mm}
  \caption{\small Pentagonal defects rotate the orientation of
    hexagons by (a) $\theta_1 = 24$$^\circ$ and (b) $\theta_2 =
    12$$^\circ$.}
  \label{fig4}
\end{figure}

Figures \ref{fig4}(a) and \ref{fig4}(b) illustrate the effect of
pentagonal defects on the orientation of hexagons at point $p^\star$ of
Figure \ref{fig3}(b). In most of the density profiles corresponding to
this coverage, we find the behavior shown in Figure \ref{fig4}(a).
Here, the effect of the pentagonal defect, which is the center of a
dislocation in the hexagonal structure, is to rotate the orientation
of the hexagons above the pentagon by $2 \cdot 60^\circ/5 = 24^\circ$
with respect to the hexagons below the pentagon.  The possible
rotations are $n \cdot 12^\circ$, where $n=$ 1,2,3,4, or 5.  The
rotation by $12^\circ$ is usually mediated by more than one equivalent
pentagon, as is shown in Figure \ref{fig4}(b) (note the ``up''
pentagon at the middle-bottom part and the ``down'' pentagon near the
middle-top part of the figure).  The ``up'' pentagon (with
one vertex on the top) is equivalent with the ``down'' pentagon
(with one vertex on the bottom) since they have five
orientationally equivalent sides).  These pentagonal defects are
induced by the fivefold symmetry of the substrate, and their
concentration decreases in the subsequent layers.

%\vspace{-3mm}

\begin{figure}[htb]
  \centerline{\epsfig{file=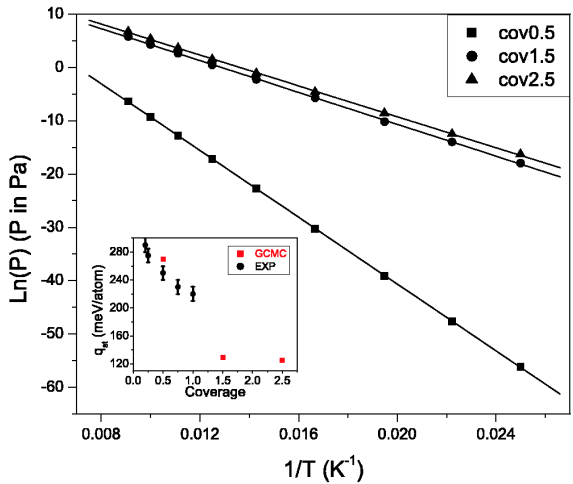,width=85mm,clip=}}
  \vspace{-3mm}
  \caption{\small (color online).
    Locations in $P$, $T$ of the vertical risers in the
    isotherms corresponding to the first (square), second (circle),
    and third (triangle) layer formation.  The heats of adsorptions,
    $q_{st}$, are 270, 129, and 125 meV/atom respectively, calculated
    as described in the text.  The inset figure shows $q_{st}$
    obtained from the simulations as well as from the experiments.}
  \label{fig5}
\end{figure}

%%%%%%%%%%%%%%% ISOSTERIC HEAT OF ADSORPTION
{\it Isosteric heat of adsorption.} Figure \ref{fig5} shows a $P$-$T$
diagram for three different coverages constructed from the isotherms
in the range 40K$<T<$110K.  In the GCMC simulations the layers grow
step-wise; at 70K the first step occurs between coverage $\sim$0.06
and $\sim$0.7, the second step occurs between coverage 1.0 and
$\sim$1.9, and the third step occurs between coverage $\sim$1.9 and
$\sim$2.8 (unit is in fractions of monolayer).  Figure \ref{fig5}
shows the $T$, $P$ location of these steps, denoted ``cov 0.5'', ``cov
1.5'', and ``cov 2.5'' for the first, second, and third steps,
respectively.  The isosteric heat of adsorption per atom at these
steps can be calculated from this $P$-$T$ diagram as
follows\cite{newref1}:
\begin{equation}
q_{st} \equiv -k_B \frac{d(ln P) }{d(1/T)_n}.
\end{equation}

%\vspace{-10mm}
The inset of Figure \ref{fig5} summarizes the values of $q_{st}$
obtained from simulations and experiments. The agreement between
experiment and the simulations for the half monolayer heat of
adsorption is good. The values obtained in the simulation for the 1.5
and 2.5 layer heats are about 20$\%$ lower than the bulk value of 165 meV
\cite{ref24}. The lower values suggest that bulk formation should be
preferred at coverages above one layer. However, layer-by-layer growth
is observed at all $T$ for at least the first few layers in these
simulations. We therefore believe that the low heats of adsorption
arise from slight inaccuracies in the Xe-Xe LJ parameters used in this
calculation, as the heats of adsorption are very sensitive to the gas
parameters.

%%%%%%%%%%%%%%% VERTICAL CELL DIMENSION
{\it Effect of the vertical dimension.} In our earlier study only 2
steps, corresponding to the first and second layer adsorption, were
apparent in the isotherms\cite{ref12}.  Further simulations indicate
that when the cell is extended in the vertical direction, additional
steps are observed.  Therefore the number of observable steps is
related to the size of the cell.  Nevertheless, layering is clearly
evident in the $\rho(z)$ profile, and the main features of the film
growth are not altered.  The average interlayer distance is calculated
to be about 0.37 nm, compared to 0.358 nm for the interlayer distance
in the $<111>$ direction of bulk Xe\cite{ref23}.  Our simulations of
multilayer films show variable adsorption as the simulation cell is
expanded in the direction perpendicular to the surface.  This is a
result of sensitivity to perturbations (here, cell size) close to the
bulk chemical potential, where the wetting film's compressibility
diverges.  This dependence has been seen previously in large scale
simulations.  See e.g. Figure 3 of reference \cite{refTRESHOLD}.  The
analog of this effect in real experiments is capillary condensation at
pressures just below saturated vapor pressure (svp), the difference
varying as the inverse pore radius.

%%%%%%%%%%%%%%%%%%%%%%%%%%%%%%%%%%%%%%%%%%%%%%%%%%%%%%%%%%%%%%%%%%%%
\section{Conclusions}

One of the main motivations for the study of Xe on the QC surface was
to elucidate which adsorption phenomena are due to the QC structure of
the substrate, as opposed to chemical interactions between the adatoms
and the substrate. We have discovered a system that is rich with
interesting phenomena, some of which are common to other physisorption
systems and some that are different. The features that have been
observed include layer-by-layer growth at low temperatures, complete
wetting above the 3D triple point, a first-order phase transition from
a ``commensurate'' structure to an incommensurate hexagonal
close-packed structure within the monolayer regime, substrate-induced
alignment of the incommensurate film, and an increase in defect
probability with temperature.  Above a monolayer, the structure
continues to grow in hexagonal close-packed layers \cite{ref12}.  The
features that are different are specific to the fivefold symmetry of
the substrate, and include a fivefold commensurate structure and a
fivefold to sixfold structural transition. We have observed some other
phenomena for which we are unaware of previous reports, including the
U-shaped curve of commensurate-incommensurate transition chemical
potential $\mu^\star_{tr}$ versus temperature and orientational domain
boundaries generated by pentagonal defects. It would be interesting to
investigate how these features are affected by different potential
parameters, and further investigations for Ne, Ar and Kr adsorption on
this surface are in progress\cite{refprc}.

The agreement between the simulations reported here and the corresponding 
experimental studies by our group are very good, perhaps better than might 
be expected, although good correspondence was already observed earlier in 
calculations of the low-coverage properties of this system\cite{ref20}. 
The main points of comparison are the nature of the film growth, 
the submonolayer isosteric heat of adsorption, and the structure 
of the film above one monolayer. The high level of agreement for 
these features indicates that LJ potentials provide a 
good description of the interactions in this system, at least those 
pertaining to the main features of the film growth. 
Despite this good agreement, one might question the use of LJ potentials 
in light of experimental observations that Xe atoms on metal surfaces have 
a preference for low-coordination sites\cite{newref3}. 
That preference is believed to originate from a screening response of the 
metal to the adsorbed Xe\cite{newref4}. At this time, there are no 
experimental measurements of adsorption sites for Xe on QC surfaces, but 
the good agreement found here for film growth suggests that any such 
screening interactions have a negligible effect on the global adsorption 
behavior, which is also the case for metal surfaces. 

It would be very useful to have experimental measurements that would 
elucidate the structure of the monolayer Xe film. LEED experiments 
sample the outer several layers of the sample, making it difficult to 
differentiate between the Xe structure and the substrate structure if 
they are the same, as indicated by these simulations. 
Low-temperature STM experiments on this system have thus far been 
inconclusive because of the difficulty of establishing a tunneling 
current through the Xe to the weakly-conducting QC substrate. 
An ideal probe of the monolayer structure would be He-atom diffraction, 
which has been used for similar measurements of metal films on 
QC surfaces\cite{newref5}, and such measurements are planned. 

%%%%%%%%%%%%%%%%%%%%%%%%%%%%%%%%%%%%%%%%%%%%%%%%%%%%%%%%%%%%%%%%%%%%
\section{Acknowledgments}
We gratefully acknowledge useful interactions with Raluca A. Trasca,
Chris Henley, David Rabson, Aleksey Kolmogorov, Katariina Pussi,
Hsin-I Li, and L. W. Bruch. We acknowledge the San Diego Supercomputer
Center for computing time under Proposal Number MSS060002. This
material is based upon work supported by the National Science
Foundation under Grant Nos. 0208520 and 0505160.
We are grateful to critical comments of the referees of this paper.

%%%%%%%%%%%%%%%%%%%%%%%%%%%%%%%%%%%%%%%%%%%%%%%%%%%%%%%%%%%%%%%%%%%%

\end{document}